\documentclass[aps,prd,reprint,twocolumn,superscriptaddress]{revtex4-1}
\usepackage{amsmath}
\usepackage{amssymb}
\usepackage{graphicx}
\usepackage{color}
\usepackage{times}
\usepackage{euscript}
\usepackage{amsthm}
\usepackage{amsfonts}
\usepackage[english]{babel}
\usepackage{epstopdf}
\usepackage{multirow,makecell,array}
\usepackage{mathtools}
\usepackage{float}
\usepackage[dvipsnames]{xcolor}
\usepackage{graphicx}

\newcommand*{\I}{ {\rm i} }

\begin{document}

\title{Pair production in rotating electric fields via quantum kinetic equations: Resolving helicity states}

\author{I.~A.~Aleksandrov}
\email{i.aleksandrov@spbu.ru}
\affiliation{Department of Physics, Saint Petersburg State University, Universitetskaya Naberezhnaya 7/9, Saint Petersburg 199034, Russia}
\affiliation{Ioffe Institute, Politekhnicheskaya street 26, Saint Petersburg 194021, Russia}
\author{A.~Kudlis}
\email{andrewkudlis@gmail.com}
\affiliation{Abrikosov Center for Theoretical Physics, MIPT, Dolgoprudnyi, Moscow Region 141701, Russia}
\affiliation{Russian Quantum Center, Skolkovo, Moscow 121205, Russia}

\begin{abstract}
We investigate the phenomenon of electron-positron pair production from vacuum in the presence of a strong electric field of circular polarization. By means of a nonperturbative approach based on the quantum kinetic equations (QKEs), we numerically calculate helicity-resolved momentum distributions of the particles produced and analyze the corresponding helicity asymmetry. It is demonstrated that the external rotating field tends to generate left-handed and right-handed particles traveling in opposite directions. Generic symmetry properties of the momentum spectra are examined analytically by means of the QKEs and also confirmed and illustrated by direct numerical computations. The helicity signatures revealed in our study are expected to provide a firmer basis for possible experimental investigations of the fundamental phenomenon of vacuum pair production in strong fields.

\end{abstract}

\maketitle

{\it Introduction.} Already within the first decade after Dirac proposed his wave equation~\cite{dirac_1928} describing the relativistic quantum dynamics of the electron $e^-$, not only was the antiparticle $e^+$ (positron) predicted~\cite{dirac_1931} and experimentally discovered~\cite{anderson}, but also it became clear that the vacuum fluctuations of the electron-positron field lead to the nonlinear effect of light-by-light scattering in the presence of a classical electric background~\cite{euler_kockel}. Furthermore, the idea that the vacuum state interacting with a strong electric field can produce electron positron pairs~\cite{sauter_1931} was later confirmed within the framework of quantum electrodynamics (QED) by means of a fully nonperturbative expression for the one-loop effective action~\cite{heisenberg_euler,schwinger_1951}. It was found that the imaginary part of the effective action, which governs the probability of the pair-production process, is proportional to $\mathrm{exp} (-\pi E_\text{c}/E_0)$, where $E_0$ is the field strength of a constant electric background and $E_\text{c} = m^2 c^3/|e\hbar| \sim 10^{16}$~V/cm is the so-called critical (Schwinger) field strength ($m$ and $e$ are the electron mass and charge, respectively). This result suggests that the corresponding Sauter-Schwinger mechanism is strongly suppressed unless $E_0$ is close to $E_\text{c}$. From the theoretical point of view, it is evident that the effect of pair production in a strong quasistatic field is nonperturbative with respect to the field amplitude $E_0$. Although this fundamental phenomenon has never been experimentally observed, the continual development of the laser technologies indicates that forming a setup that involves sufficiently intense laser fields may become feasible in the near future (for review, see Ref.~\cite{fedotov_review}).

There are several theoretical techniques which can be employed for describing the nonperturbative pair-production effect (see, e.g., Refs.~\cite{fradkin_gitman_shvartsman,BB_prd_1991,gavrilov_prd_1996,zhuang_1996,zhuang_prd_1998,ochs_1998,schmidt_1998,kluger_prd_1998,pervushin_skokov,hebenstreit_prd_2010,blaschke_prd_2011,blinne_gies_2014,woellert_prd_2015,blinne_strobel_2016,aleksandrov_prd_2016,li_prd_2017,lv_pra_2018,aleksandrov_epjst_2020,aleksandrov_kohlfuerst,aleksandrov_kudlis_klochai} and references therein). Here we will utilize the system of quantum kinetic equations (QKEs), which provides an exact treatment of vacuum pair production in spatially uniform electric backgrounds of arbitrary polarization. In Ref.~\cite{aleksandrov_kudlis_klochai} it was shown that the system involves ten unknown functions and differs from the QKEs obtained previously~\cite{pervushin_skokov} due to an inaccuracy in the preceding derivations. Here we will perform numerical computations based on the correct form of the QKEs.

Our main focus is placed on the spin effects, which manifest themselves in setups involving rotating electric fields. Since the number densities of the particles produced are defined as functions of momentum, the spin projection onto a given laboratory-frame axis is not a conserved quantum number as the corresponding spin operator does not commute with the one-particle Hamiltonian, and it is the {\it total} angular momentum that represents a conserved quantity. Here we propose an experimentally relevant strategy based on measuring the particle’s helicity, i.e. spin projection onto the propagation direction. The corresponding helicity-resolved densities can be obtained from the QKE formalism~\cite{aleksandrov_kudlis_klochai}.
Our goal is to describe the helicity asymmetry of the electrons and positrons produced in the presence of a rotating external background, which mimics the QED dynamics in a combination of two counterpropagating laser pulses of circular polarization. The momentum distributions of left- and right-handed particles represent evident observable signatures that should facilitate the experimental efforts concerning practical investigations of the Sauter-Schwinger effect.

Throughout this letter, we assume $\hbar = c = 1$ and use $e<0$ as the electric charge of the electron $e^-$.

\mbox{}

{\it Theoretical framework.} To describe the external classical background, we employ the temporal gauge $A_0=0$, $\mathbf{A} = \mathbf{A} (t)$. The corresponding electric field $\mathbf{E} (t) = -\dot{\mathbf{A}} (t)$ is assumed to vanish for $t \leqslant t_\text{in}$ and $t \geqslant t_\text{out}$. The electron-positron field is quantized in the presence of the external field, i.e. within the Furry picture. Since the electromagnetic background is homogeneous in space, the Heisenberg field operator can be represented by means of the {\it adiabatic} Hamiltonian eigenfunctions in the following form:
\begin{eqnarray}
\psi (x) &=& \sum_{s=\pm 1} \int \! \frac{d\mathbf{p}}{(2\pi)^{3/2}} \, \mathrm{e}^{\I \mathbf{p} \mathbf{x}}\big [a_{\mathbf{p},s} (t) u_{\mathbf{p} - e \mathbf{A}(t), s} (t) \nonumber \\
{}&+&b^\dagger_{-\mathbf{p},s} (t) v_{\mathbf{p} - e \mathbf{A}(t), s} (t) \big ] \,. \label{eq:psi_adiabatic}
\end{eqnarray}
Here $x = (t, \, \mathbf{x})$ and the bispinors are chosen in the form
\begin{eqnarray}
u_{\boldsymbol{p},s} &=& C(p^0)
\begin{pmatrix}
(p^0 + m) w_s\\
(\boldsymbol{\sigma} \mathbf{p}) w_s
\end{pmatrix} \,, \label{eq:u_explicit}\\
v_{\boldsymbol{p},s} &=& C(p^0)
\begin{pmatrix}
-(\boldsymbol{\sigma} \mathbf{p}) w_s\\
(p^0 + m) w_s
\end{pmatrix} \,,\label{eq:v_explicit}
\end{eqnarray}
where $C(p^0) = [2p^0 (p^0 +m)]^{-1/2}$, $p^0 = p_0 = \sqrt{m^2 + \boldsymbol{p}^2}$, and $w_{-1} = (1, \, 0)^\mathrm{t}$, $w_{+1} = (0, \, 1)^\mathrm{t}$ \,.

Let us denote the Heisenberg state corresponding to the initial vacuum by $| 0,\text{in} \rangle$. It differs from the final vacuum state $| 0,\text{out} \rangle$ due to the vacuum instability~\cite{fradkin_gitman_shvartsman}. The real-valued QKE functions can be introduced by the following expresssions for the in-vacuum expectation values~\cite{aleksandrov_kudlis_klochai}:
\begin{widetext}
\begin{eqnarray}
\langle 0, \text{in} | a^\dagger_{\mathbf{p},s} (t) a_{\mathbf{p}',s'} (t) | 0,\text{in} \rangle &=& \delta (\mathbf{p} - \mathbf{p}') \big [ f (\mathbf{p}, t) \delta_{s's} + \mathbf{f} (\mathbf{p}, t) \boldsymbol{\sigma}_{s's} \big ] \,, \label{eq:in_mean_qke_ad_a} \\
\langle 0, \text{in} | b^\dagger_{\mathbf{p},s} (t) b_{\mathbf{p}',s'} (t) | 0,\text{in} \rangle &=& \delta (\mathbf{p} - \mathbf{p}') \big [ f (-\mathbf{p}, t) \delta_{ss'} - \mathbf{f} (-\mathbf{p}, t) \boldsymbol{\sigma}_{ss'} \big ] \,, \label{eq:in_mean_qke_bd_b} \\
\langle 0, \text{in} | a^\dagger_{\mathbf{p},s} (t) b^\dagger_{\mathbf{p}',s'} (t) | 0,\text{in} \rangle &=& \delta (\mathbf{p} + \mathbf{p}') \big \{ [\mathbf{u} (\mathbf{p}, t) - \I \mathbf{v} (\mathbf{p}, t) ] \boldsymbol{\sigma}_{s's} \big \} \,, \\
\langle 0, \text{in} | b_{\mathbf{p},s} (t) a_{\mathbf{p}',s'} (t) | 0,\text{in} \rangle &=& \delta (\mathbf{p} + \mathbf{p}') \big \{ [\mathbf{u} (-\mathbf{p}, t) + \I \mathbf{v} (-\mathbf{p}, t) ] \boldsymbol{\sigma}_{s's} \big \} \,.
\end{eqnarray}
\end{widetext}
By utilizing the equation of motion for the Heisenberg field operator, one can formulate the problem in terms of the QKE system governing the temporal evolution of the ten unknown components $f$, $\mathbf{f}$, $\mathbf{u}$, and $\mathbf{v}$ (a detailed derivation is presented in Ref.~\cite{aleksandrov_kudlis_klochai}):
\begin{eqnarray}
\dot{f} &=& -2 \boldsymbol{\mu}_2 \mathbf{u} \,, \label{eq:system_simple_f_s}\\
\dot{\mathbf{f}} &=& 2 (\boldsymbol{\mu}_1 \times \mathbf{f}) - 2 (\boldsymbol{\mu}_2 \times \mathbf{v}) \,, \label{eq:system_simple_f_v} \\
\dot{\mathbf{u}} &=& 2 (\boldsymbol{\mu}_1 \times \mathbf{u}) + \boldsymbol{\mu}_2 (2f - 1) + 2 \omega \mathbf{v} \,, \label{eq:system_simple_u_v} \\
\dot{\mathbf{v}} &=& 2 (\boldsymbol{\mu}_1 \times \mathbf{v}) - 2 (\boldsymbol{\mu}_2 \times \mathbf{f}) - 2 \omega \mathbf{u} \,. \label{eq:system_simple_v_v}
\end{eqnarray}
Here
\begin{eqnarray*}
\boldsymbol{\mu}_1 (\mathbf{p}, t) &=& \frac{e}{2\omega (\omega+m)}\, [\mathbf{q} \times \mathbf{E} (t)] \,, \label{eq:mu_1} \\
\boldsymbol{\mu}_2 (\mathbf{p}, t) &=& \frac{e}{2\omega^2 (\omega+m)}\, \big \{ [\mathbf{q} \mathbf{E}(t)] \mathbf{q} - \omega (\omega+m) \mathbf{E} (t) \big \} \,, \label{eq:mu_2} 
\end{eqnarray*}
and $\mathbf{q} \equiv \mathbf{p}-e\mathbf{A} (t)$, $\omega \equiv \sqrt{m^2 + \mathbf{q}^2}$. All of the QKE functions vanish at $t = t_\text{in}$. While they depend on $\mathbf{p}$, it is $\mathbf{q}$ that corresponds to the kinetic (gauge-invariant) momentum of the particles, so the final values of the QKE components will be evaluated as a function of $\mathbf{q} (t_\text{out}) = \mathbf{p}-e\mathbf{A} (t_\text{out})$. We underline that the vector potential does not necessarily vanish as $t \to \infty$.

The system~\eqref{eq:system_simple_f_s}--\eqref{eq:system_simple_v_v} differs from the QKEs obtained almost two decades ago~\cite{pervushin_skokov}. In Ref.~\cite{aleksandrov_kudlis_klochai} it was demonstrated that the new (correct) form of the QKEs is completely equivalent to the system that appears within the Dirac-Heisenberg-Wigner formalism~\cite{BB_prd_1991,zhuang_1996,zhuang_prd_1998,ochs_1998,hebenstreit_prd_2010,blinne_gies_2014,blinne_strobel_2016,li_prd_2017,lv_pra_2018,aleksandrov_kohlfuerst,aleksandrov_kudlis_klochai,olugh_prd_2019,kohlfuerst_prd_2019,kohlfuerst_arxiv_2022_2,yu_prd_2023,hu_prd_2023,hu_arxiv_2024}.

In what follows, we will solve the system~\eqref{eq:system_simple_f_s}--\eqref{eq:system_simple_v_v} numerically and extract the necessary helicity-resolved densities at $t = t_\text{out}$.

\mbox{}

{\it Helicity states.} Although the QKE functions evolved up to $t=t_\text{out}$ can already be used to extract the spin-resolved densities via Eqs.~\eqref{eq:in_mean_qke_ad_a} and \eqref{eq:in_mean_qke_bd_b}, the ``spin quantum number'' $s$ does not provide a physically relevant quantity as it allows one to merely obtain the populations of the states corresponding to the {\it basis} bispinors~\eqref{eq:u_explicit} and \eqref{eq:v_explicit}. These are not the eigenfunctions of the spin-projection operator. Moreover, such eigenfunctions cannot be constructed once one fixes the generalized (canonical) momentum $\mathbf{p}$ since the spin operator does not commute with the one-particle Hamiltonian. We propose to unitary transform the basis bispinors, so that they become the eigenfunctions of the {\it helicity operator} $(\boldsymbol{\Sigma} \mathbf{p})/|\mathbf{p}|$, which does commute with the Hamiltonian. Accordingly, the new states possess a well-defined quantum number in addition to $\mathbf{p}$. Note that in order to preserve a relatively simple form of the QKEs~\eqref{eq:system_simple_f_s}--\eqref{eq:system_simple_v_v}, we will keep the same kinetic functions and employ them at the very end of the temporal propagation to construct the helicity-resolved momentum densities. As was shown in Ref.~\cite{aleksandrov_kudlis_klochai}, in the case of electrons, the latter read
\begin{equation}
f^{(e^-\text{L/R})}(\mathbf{p}, t) = f(\mathbf{p}, t) \mp \frac{\mathbf{q} \mathbf{f} (\mathbf{p}, t)}{|\mathbf{q}|} \,. \label{eq:el_LR}
\end{equation}
Here L (R) corresponds to the left-handed (right-handed) particle, i.e. to negative (positive) helicity. The positron distributions can be then easily obtained via
\begin{equation}
f^{(e^+\text{L/R})}(\mathbf{p}, t) = f^{(e^-\text{R/L})}(-\mathbf{p}, t) \,.
\label{eq:pos_LR}
\end{equation}
The final particle distributions are those calculated at $t=t_\text{out}$. For instance,
\begin{equation}
\frac{(2\pi)^3}{V} \frac{dN^{(e^-\text{L})}_{\mathbf{q}}}{d^3 \mathbf{q}} = f^{(e^-\text{L})}(\mathbf{q} + e \mathbf{A} (t_\text{out}), t_\text{out}) \,,
\end{equation}
where $V$ is the volume of the system and $\mathbf{q}$ is the particle's kinetic momentum. Note that due to the Pauli exclusion principle, these densities never exceed unity as can also be directly proved by inspecting the system~\eqref{eq:system_simple_f_s}--\eqref{eq:system_simple_v_v}~\cite{aleksandrov_kudlis_klochai}.

\mbox{}

{\it Numerical results.} The rotating external field is chosen in the following form:
\begin{equation}
\mathbf{E} (t) = \frac{E_0}{\sqrt{2}}  \, F(\Omega t) \big [ \cos (\Omega t) \, \mathbf{e}_x + \sin (\Omega t) \, \mathbf{e}_y \big ],
\label{eq:field}
\end{equation}
where $\mathbf{e}_x$ ($\mathbf{e}_y$) is the Cartesian unit vector along the $x$ ($y$) direction, $F(\eta) = \mathrm{exp}(-\eta^2/\sigma^2)$ is a pulse envelope, and $\sigma$ is a dimensionless parameter governing the pulse duration. We set $t_\text{out} = -t_\text{in} = T$, where $T$ is sufficiently large, so the external field is practically zero for $|t|>T$ and the numerical results are converged. The setup~\eqref{eq:field} approximates a combination of two counterpropagating circularly-polarized laser pulses of opposite handedness (the role of the laser handedness was explored in terms of the number density $f$ in Ref.~\cite{kohlfuerst_arxiv_2022_2}).

\begin{figure}[t]
    \centering
    \includegraphics[width=0.98\linewidth]{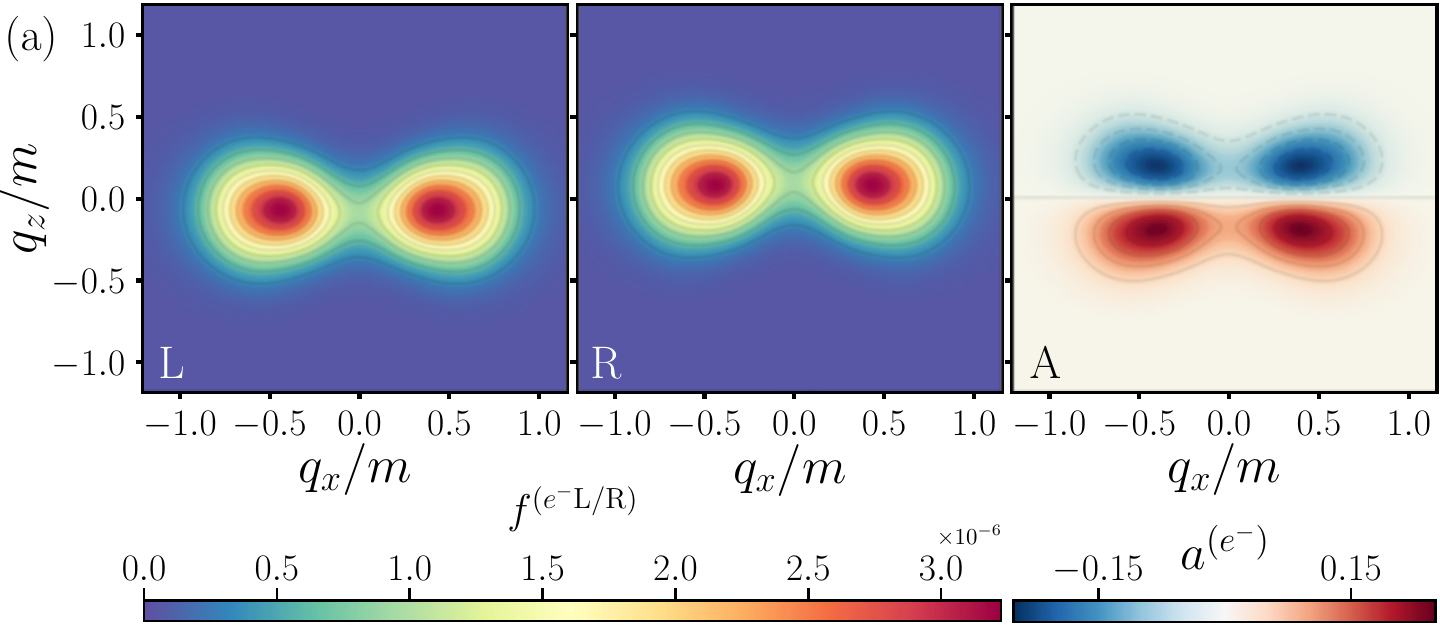}\\
  \includegraphics[width=0.98\linewidth]{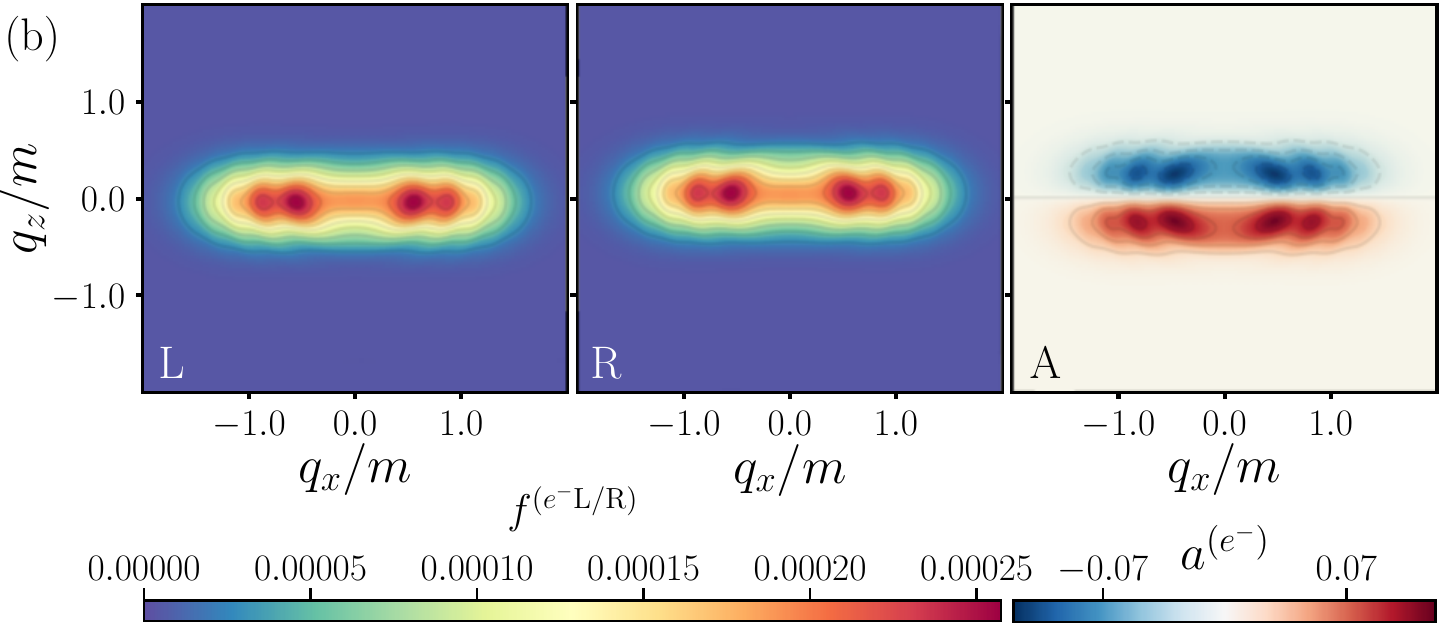}\\
  \includegraphics[width=0.98\linewidth]{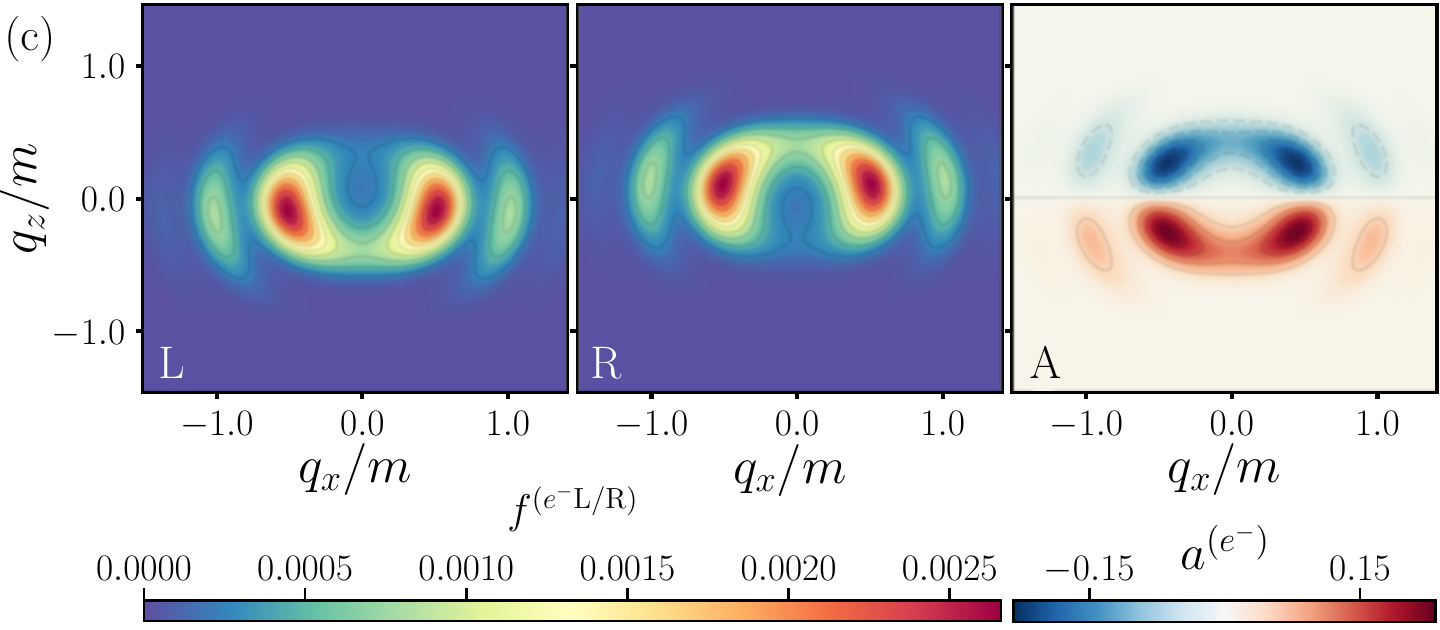}
    \caption{Helicity-resolved momentum distributions of the electrons produced by the electric field~\eqref{eq:field} as a function of the kinetic momentum projections $q_x$ and $q_z$. Left and middle: number densities $f^{(e^-\text{L/R})}$. The right panels correspond to the helicity asymmetry. The field parameters are (a) $E_0 = 0.2E_\text{c}$, $\Omega = 0.5 m$, $\sigma = 5$; (b) $E_0 = 0.5E_\text{c}$, $\Omega = 0.2 m$, $\sigma = 10$; (c) $E_0 = 0.5E_\text{c}$, $\Omega = 0.5 m$, $\sigma = 10$. The spectra were obtained for $p_y = 0$ in the gauge $\mathbf{A}(0)=0$.}
    \label{fig:spectra_xz}
\end{figure}

Before we present the results of our numerical calculations, let us discuss a fundamental symmetry property of the QKE system~\eqref{eq:system_simple_f_s}--\eqref{eq:system_simple_v_v} which can be established analytically. One can directly verify that changing the sign of $p_z = q_z$, one obtains a solution with opposite signs of $f_x$, $f_y$, $u_z$, and $v_z$ (the other components remain the same). In this case, the product $\mathbf{q} \mathbf{f}/|\mathbf{q}|$ changes the sign, so the helicity-resolved distributions ``L’’ and ``R’’ turn to each other upon the reflection $q_z \to -q_z$ (or, equivalently, $p_z \to -p_z$). In other words, the first (second) term in Eq.~\eqref{eq:el_LR} is symmetric (antisymmetric) with respect to $q_z \to - q_z$. There are several very important consequences of this symmetry. First, the helicity asymmetry $f^{(e^-\text{L})} - f^{(e^-\text{R})}$, which immediately appears once $\mathbf{q} \mathbf{f}/|\mathbf{q}| \neq 0$, is an asymmetric function of $q_z$. Note that in the case of a linearly polarized external field, the function $\mathbf{f}$ vanishes identically~\cite{aleksandrov_kudlis_klochai}. Second, it is sufficient to numerically calculate only one of the two partial distributions [either $f^{(e^-\text{L})}(\mathbf{p}, t)$, or $f^{(e^-\text{R})}(\mathbf{p}, t)$], and the helicity asymmetry is determined by the $q_z$ asymmetry of the corresponding spectra. Finally, we note that these features take place once $E_z = 0$, so they hold within a broad class of the field configurations (e.g., the field can have arbitrary elliptical polarization). The total number density summed over helicity is always symmetric with respect to $q_z \to -q_z$.

Since the helicity asymmetry vanishes for $p_z=q_z=0$, in what follows we will focus on the analysis of $q_x q_z$ and $q_y q_z$ momentum distributions. Even the total number densities $f^{(e^-)}$, i.e. those summed over the spin states, being computed within these momentum planes are not frequently discussed in the literature ($q_xq_y$ spectra were examined in numerous studies~\cite{blinne_gies_2014,blinne_strobel_2016,li_prd_2017,olugh_prd_2019,hu_prd_2023,hu_arxiv_2024}). In Fig.~\ref{fig:spectra_xz} we present the typical $q_x q_z$ momentum spectra for three different sets of the field parameters. The left and middle graphs correspond to negative and positive helicity, respectively, while in the right column we display the degree $a^{(e^-)}$ of the helicity asymmetry defined as the difference $f^{(e^-\text{L})} - f^{(e^-\text{R})}$ divided by the maximal value of the electron number density within the corresponding momentum plane. We observe that this quantity is indeed an odd function of $q_z$ and its sign is opposite to that of $q_z$. This indicates that the right-handed electrons preferably move along the $z$ axis, whereas the left-handed particles tend to travel in the opposite direction. Furthermore, due to the property~\eqref{eq:pos_LR}, the asymmetry degree of the positrons produced is also opposite to the sign of the $z$ momentum projection, so the right-handed positrons are also likely to propagate in the positive direction of the $z$ axis. The asymmetry degree in our examples is of the order of $10$--$20\%$, which represents a macroscopic value from the experimental viewpoint.

Another important feature of the momentum distributions is its symmetry with respect to the reflection $p_x \to -p_x$. As was shown in Refs.~\cite{blinne_gies_2014,dumlu_dunne_prd_2011}, this property appears due to the time-reversal symmetry. In terms of the QKE components, we find that the transformation $p_x \to -p_x$, $t \to -t$ does not alter the QKEs~\eqref{eq:system_simple_f_s}--\eqref{eq:system_simple_v_v}, provided $f_x$, $u_x$, $v_y$, and $v_z$ change the sign. The crucial point here is that since in this case $q_x \to -q_x$, the second term in Eq.~\eqref{eq:el_LR} remains the same, so the symmetry also holds for the {\it helicity-resolved} distributions. Note that it is essential that the $x$ ($y$) component in Eq.~\eqref{eq:field} is symmetric (antisymmetric) under time reversal, and the generalized momentum $p_x$ corresponds to the gauge $\mathbf{A} (0) = 0$, so the vector potential has also symmetry properties. For the field parameters chosen in our examples, the difference between $q_x$ and $p_x$ can hardly be identified in the graphs. One of the important consequences of the $p_x$ symmetry is that it completely prohibits any vortex structures of the helicity-resolved densities involving the $p_x$ axis (cf. Ref.~\cite{hu_arxiv_2024}, where instead of the helicity-resolved spectra, more complex distributions were analyzed). We also note that the total density, i.e. the sum of the left and middle plots, computed by means of our numerical procedures exhibits the usual behavior reported, e.g., in Refs.~\cite{blinne_strobel_2016,hu_arxiv_2024}.

\begin{figure}[t]
    \centering
  \includegraphics[width=0.98\linewidth]{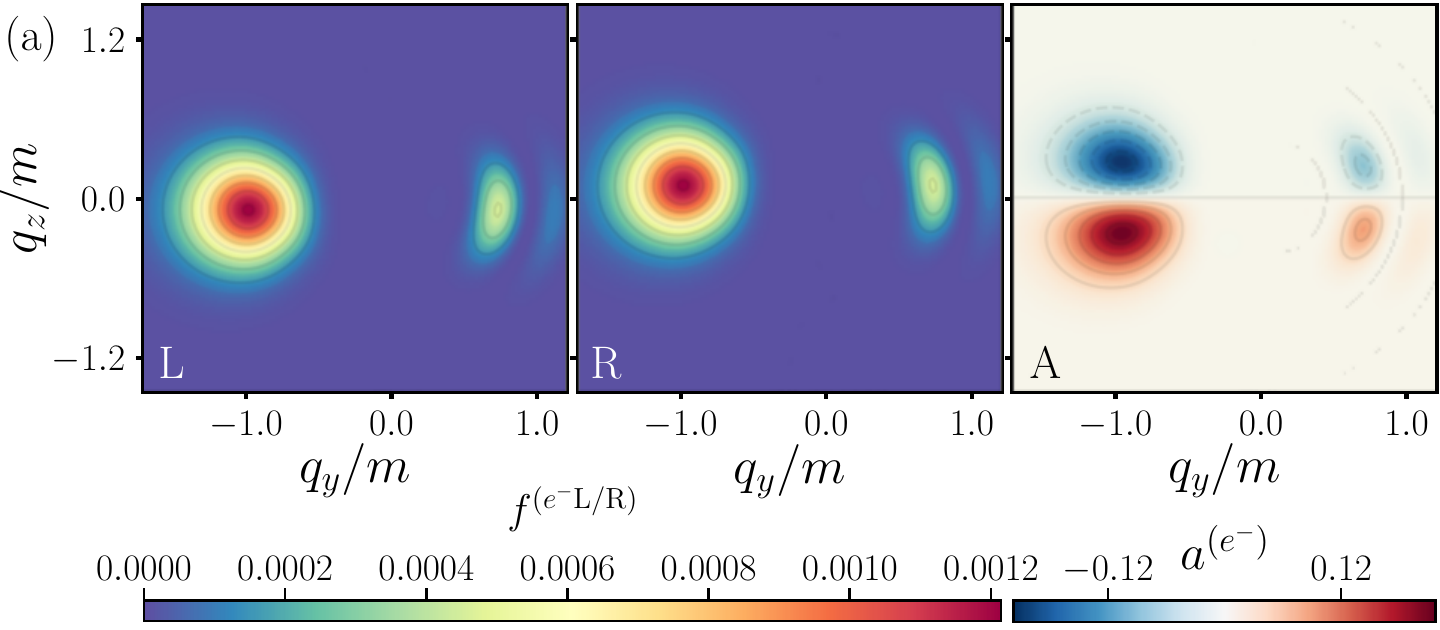}\\
  \includegraphics[width=0.98\linewidth]{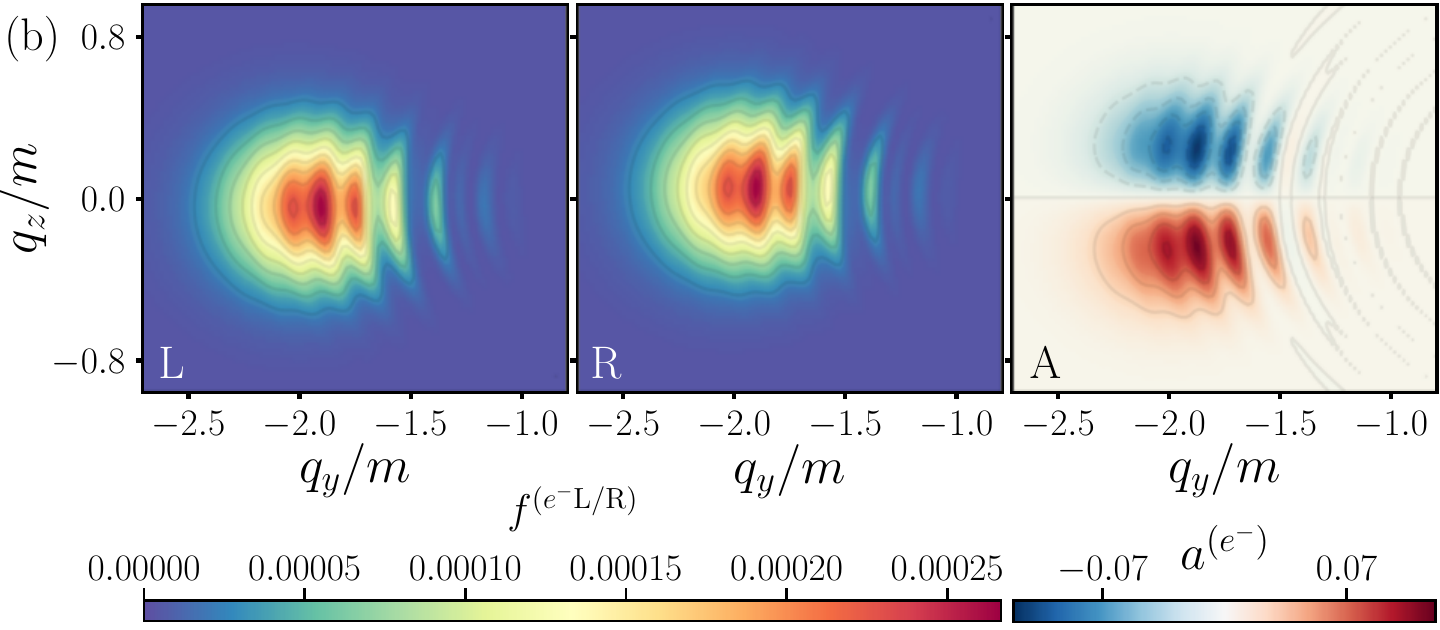}
    \caption{Helicity-resolved momentum distributions of the electrons as a function of $q_y$ and $q_z$. Left and middle: number densities $f^{(e^-\text{L/R})}$. The right panels correspond to the helicity asymmetry. The field parameters are (a) $E_0 = 0.5E_\text{c}$, $\Omega = 0.5 m$, $\sigma = 5$; (b) $E_0 = 0.5E_\text{c}$, $\Omega = 0.2 m$, $\sigma = 10$. The spectra were obtained for $p_x = 0$ in the gauge $\mathbf{A}(0)=0$.}
    \label{fig:spectra_yz}
\end{figure}

In Fig.~\ref{fig:spectra_yz} we depict analogous momentum distributions in the $q_yq_z$ plane. Here we observe again the $q_z$ symmetry, while the spectra are not symmetric with respect to $q_y$. Note that in Fig.~\ref{fig:spectra_yz}(b) the electron distributions are considerably shifted in the region of negative $q_y$. This feature appears due to a great net momentum transfer in this direction because of the large value of the corresponding electric-field area, i.e. large value of $A_y (t_\text{out})$. This effect with respect to the $x$ axis (Fig.~\ref{fig:spectra_xz}) is not evident as was pointed out above. Note also that in Fig.~\ref{fig:spectra_xz} we used $p_y=0$, which yields quite large values of the electron densities unlike, e.g., $q_y=0$. Since the external field does not exert a force in the $z$ direction, there is no net acceleration with respect to $q_z$ and the support of the momentum distributions corresponds to $q_z \lesssim m$. Nevertheless, the helicity asymmetry is obvious: the laser pulse tends to push right-handed electrons toward the $z$ direction, while the left-handed electrons are mainly accelerated in the opposite direction. In terms of measuring the scattering angles, this asymmetry should be evidently identified as the plane $q_z = 0$ clearly divides the space into two different subspaces. The generic symmetry properties uncovered above and our numerical examples demonstrate rather universal patterns. We also verified that if one changes the handedness of the external rotating field by reversing the sign of the second term in Eq.~\eqref{eq:field}, then the momentum distributions will yield the opposite sign of the helicity asymmetry. Finally, we note that in Fig.~\ref{fig:spectra_yz}(b) the momentum distribution has a pronounced interference structure, which is a well-elaborated effect frequently occurring in the case of many-cycle pulses~\cite{blinne_gies_2014,blinne_strobel_2016,li_prd_2017,olugh_prd_2019,hu_prd_2023,hu_arxiv_2024,dumlu_dunne_prd_2011} [see also Fig.~\ref{fig:spectra_xz}(b) and (c), where $\sigma = 10$].

\mbox{}

{\it Conclusions.} In this study, we numerically implemented a recently derived system of quantum kinetic equations and investigated the process of nonperturbative vacuum pair production in a rotating electric background. To address the spin effects, we focused on the analysis of the particle's helicity, which represents a well-defined quantity from the quantum mechanical perspective. By inspecting the helicity-resolved distributions of the electrons and positrons, we revealed a significant helicity asymmetry providing a distinct observable feature. Our findings should further illuminate the physical intricacies of vacuum pair production and facilitate experimental research concerning strong-field QED.

\mbox{}

{\it Acknowledgments.} The investigation was funded by the Ministry of Science and Higher Education of the Russian Federation (Goszadaniye), project No. 2019-1246. I.A.A. also acknowledges the support from the Foundation for the advancement of theoretical physics and mathematics ``BASIS''.



\begin{thebibliography}{99}
%
\bibitem{dirac_1928} P.~A.~M. Dirac, The Quantum theory of electron, Proc. Roy. Soc. Lond. {\bf A117}, 610 (1928).
%
\bibitem{dirac_1931} P.~A.~M. Dirac, Quantized singularities in the electromagnetic field, Proc. Roy. Soc. Lond. {\bf A133}, 60 (1931).
%
\bibitem{anderson} C.~D.~Anderson, The positive electron, Phys. Rev. {\bf 43}, 491 (1933).
%
\bibitem{euler_kockel} H.~Euler and B.~Kockel, The scattering of light by light in the Dirac theory, Naturwiss. {\bf 23}, 246 (1935).
%
\bibitem{sauter_1931} F.~Sauter, \"Uber das Verhalten eines Elektrons im homogenen elektrischen Feld nach der relativistischen Theorie Diracs, Z.~Phys. {\bf 69}, 742 (1931).
%
\bibitem{heisenberg_euler} W.~Heisenberg and H.~Euler, Folgerungen aus der Diracschen Theorie des Positrons, Z.~Phys. {\bf 98}, 714 (1936).
%
\bibitem{schwinger_1951} J.~Schwinger, On gauge invariance and vacuum polarization, Phys. Rev. {\bf 82}, 664 (1951).
%
\bibitem{fedotov_review} A.~Fedotov, A.~Ilderton, F.~Karbstein, B.~King, D.~Seipt, H.~Taya, and G.~Torgrimsson, Advances in QED with intense background fields, Phys. Rep. {\bf 1010}, 1 (2023).
%
\bibitem{fradkin_gitman_shvartsman} E.~S.~Fradkin, D.~M.~Gitman, and S.~M.~Shvartsman, {\it Quantum electrodynamics with unstable vacuum} (Springer-Verlag, Berlin, 1991).
%
\bibitem{BB_prd_1991} I.~Bialynicki-Birula, P.~G\'ornicki, and J.~Rafelski, Phase-space structure of the Dirac vacuum, Phys. Rev. D {\bf 44}, 1825 (1991).
%
\bibitem{gavrilov_prd_1996} S.~P.~Gavrilov and D.~M.~Gitman, Vacuum instability in external fields, Phys. Rev. D {\bf 53}, 7162 (1996).
%
\bibitem{zhuang_1996} P.~Zhuang and U.~Heinz, Relativistic quantum transport theory for electrodynamics, Ann. Phys. {\bf 245}, 311 (1996).
%
\bibitem{zhuang_prd_1998} P.~Zhuang and U.~Heinz, Equal-time hierarchies for quantum transport theory, Phys. Rev. D {\bf 57}, 6525 (1998).
%
\bibitem{ochs_1998} S.~Ochs and U.~Heinz, Wigner functions in covariant and single-time formulations, Ann. Phys. {\bf 266}, 351 (1998).
%
\bibitem{schmidt_1998} S.~Schmidt, D.~Blaschke, G.~R\"opke, S.~A.~Smolyansky, A.~V.~Prozorkevich, and V.~D.~Toneev, A quantum kinetic equation for particle production in the Schwinger mechanism, Int. J. Mod. Phys. E {\bf 07}, 709 (1998).
%
\bibitem{kluger_prd_1998} Y.~Kluger, E.~Mottola, and J.~M.~Eisenberg, Quantum Vlasov equation and its Markov limit, Phys. Rev. D {\bf 58}, 125015 (1998).
%
\bibitem{pervushin_skokov} V.~N.~Pervushin and V.~V.~Skokov, Kinetic description of fermion production in the oscillator representation, Acta Phys. Polon. B {\bf 37}, 2587 (2006).
%
\bibitem{hebenstreit_prd_2010} F.~Hebenstreit, R.~Alkofer, and H.~Gies, Schwinger pair production in space- and time-dependent electric fields: Relating the Wigner formalism to quantum kinetic theory, Phys. Rev. D {\bf 82}, 105026 (2010).
%
\bibitem{blaschke_prd_2011} D.~B.~Blaschke, V.~V.~Dmitriev, G.~R\"opke, and S.~A.~Smolyansky, BBGKY kinetic approach for an $e^-e^+\gamma$ plasma created from the vacuum in a strong laser-generated electric field: The one-photon annihilation channel, Phys. Rev. D {\bf 84}, 085028 (2011).
%
\bibitem{blinne_gies_2014} A.~Blinne and H.~Gies, Pair production in rotating electric fields, Phys. Rev. D {\bf 89}, 085001 (2014).
%
\bibitem{woellert_prd_2015} A.~W\"ollert, H.~Bauke, and C.~H.~Keitel, Spin polarized electron-positron pair production via elliptical polarized laser fields, Phys. Rev. D {\bf 91}, 125026 (2015).
%
\bibitem{blinne_strobel_2016} A.~Blinne and E.~Strobel, Evolution of chirality in a multiphoton pair production process, Phys. Rev. D {\bf 93}, 025014 (2016).
%
\bibitem{aleksandrov_prd_2016} I.~A.~Aleksandrov, G.~Plunien, and V.~M.~Shabaev, Electron-positron pair production in external electric fields varying both in space and time, Phys. Rev. D {\bf 94}, 065024 (2016).
%
\bibitem{li_prd_2017} Z.~L.~Li, Y.~J.~Li, and B.~S.~Xie, Momentum vortices on pairs production by two counter-rotating fields, Phys. Rev. D {\bf 96}, 076010 (2017).
%
\bibitem{lv_pra_2018} Q.~Z.~Lv, S.~Dong, Y.~T.~Li, Z.~M.~Sheng, Q.~Su, and R.~Grobe, Role of the spatial inhomogeneity on the laser-induced vacuum decay, Phys. Rev. A {\bf 97}, 022515 (2018).
%
\bibitem{aleksandrov_epjst_2020} I.~A.~Aleksandrov, V.~V.~Dmitriev, D.~G.~Sevostyanov, and S.~A.~Smolyansky, Kinetic description of vacuum $e^+e^-$ production in strong electric fields of arbitrary polarization, Eur. Phys. J. Spec. Top. {\bf 229}, 3469 (2020).
%
\bibitem{aleksandrov_kohlfuerst} I.~A.~Aleksandrov and C.~Kohlf\"urst, Pair production in temporally and spatially oscillating fields, Phys. Rev. D {\bf 101}, 096009 (2020).
%
\bibitem{aleksandrov_kudlis_klochai} I.~A.~Aleksandrov, A.~Kudlis, and A.~I.~Klochai, Kinetic theory of vacuum pair production in uniform electric fields revisited, arXiv:2403:17204.
%
\bibitem{olugh_prd_2019} O.~Olugh, Z.~L.~Li, B.~S.~Xie, and R.~Alkofer, Pair production in differently polarized electric fields with frequency chirps, Phys. Rev. D {\bf 99}, 036003 (2019).
%
\bibitem{kohlfuerst_prd_2019} C.~Kohlf\"urst, Spin states in multiphoton pair production for circularly polarized light, Phys. Rev. D {\bf 99}, 096017 (2019).
%
\bibitem{kohlfuerst_arxiv_2022_2} C.~Kohlf\"urst, Pair production in circularly polarized waves, arXiv:2212.03180.
%
\bibitem{yu_prd_2023} C.~Yu, Evolution of chirality in a multiphoton pair production process, Phys. Rev. D {\bf 108}, 116009 (2023).
%
\bibitem{hu_prd_2023} L.~N.~Hu, O.~Amat, L.~Wang, A.~Sawut, H.~H.~Fan, and B.~S.~Xie, Momentum spirals in multiphoton pair production revisited, Phys. Rev. D {\bf 107}, 116010 (2023).
%
\bibitem{hu_arxiv_2024} L.~N.~Hu, H.~H.~Fan, O.~Amat, S.~Tang, and B.~S.~Xie, Spin effect induced momentum spiral and asymmetry degree in pair production, arXiv:2402.16476.
%
\bibitem{dumlu_dunne_prd_2011} C.~K.~Dumlu and G.~V.~Dunne, Interference effects in Schwinger vacuum pair production for time-dependent laser pulses, Phys. Rev. D {\bf 83}, 065028 (2011).

\end{thebibliography}
\end{document}